\documentclass[twocolumn,showpacs,preprintnumbers,draftclsnofoot]{revtex4}
\usepackage{amsmath}
\usepackage{amssymb}
\usepackage{graphicx}
\usepackage{dcolumn}
\usepackage{bm}
\usepackage{amssymb}
\usepackage{graphicx}
\usepackage{dcolumn}
\usepackage{bm}
\begin{document}

\title{ Goos-H$\ddot{a}$nchen shift of normally incident beam on magneto-optical meta-grating }
\author{Ma Luo\footnote{Corresponding author:swym231@163.com} }
\affiliation{School of Optoelectronic Engineering, Guangdong Polytechnic Normal University, Guangzhou 510665, China}

\begin{abstract}

Goos-H$\ddot{a}$nchen shift of normally incident optical beam to meta-grating consisted of magneto-optical material are studied. The meta-grating is consisted of compound grating of magneto-optical rod on top of a dielectric slab, which induces zone-folding quasi-bound states in the continuum at the $\Gamma$ point with non-zero group velocity. By applying stationary-phase theory, the Goos-H$\ddot{a}$nchen shift of normally incident quasi-plane wave are found to be enhanced at the reflection peak. Simulations of normally incident Gaussian beam with finite beam width show that the Goos-H$\ddot{a}$nchen shift is dependent on the beam width.

\end{abstract}

\maketitle

\section{Introduction}

The Goos-H$\ddot{a}$nchen (GH) shift in optical gratings arises from the excitation of resonant modes, which extract energy from the incident optical beam and propagate laterally before re-radiating the optical energy back into the surrounding medium \cite{GHshiftIntro}. The GH shift was firstly observed in 1947 \cite{firstGH47}. Resonant modes with high quality (Q) factors can significantly amplify the GH shift, such as interface with Brewster effects \cite{Brewster1,Brewster2,Brewster3}, surface plasmon polaritons \cite{sppGH1,sppGH2,sppGH3,sppGH4,sppGH5}, Fabry-Perot resonances \cite{fpcavityGH1,fpcavityGH2,fpcavityGH3,fpcavityGH4}, Bloch surface waves \cite{blochGH1,blochGH2}, Tamm plasmon polaritons \cite{tamnGH1,tamnGH2,tamnGH3}, and quasi-bound states in the continuum (quasi-BICs) \cite{bicGH1,bicGH2,bicGH3,bicGH4}. For instance, quasi-BICs characterized by ultra-high quality factors (Q factors) can induce extraordinarily large GH shifts with magnitudes exceeding $10^{3}\lambda$ (where $\lambda$ denotes the wavelength of the incident field) \cite{bicGH1}. Such substantial GH shifts can be applied in diverse fields by leveraging their sensitivity to structural perturbations and ability to manipulate light-matter interactions at the nanoscale, including optical sensing \cite{GHapplSensing1,GHapplSensing2}, wavelength division (de)multiplexing \cite{GHapplMulti}, and optical energy storage \cite{GHapplStore1,GHapplStore2,GHapplStore3}.

Meta-grating with sub-wavelength structure can induce BIC, which is a type of optical resonant mode above the light cone with infinite quality factor \cite{bicItSelf1,bicItSelf2,bicItSelf3,bicItSelf4}. The symmetric BIC is due to the symmetry mismatch between the BIC and the radiative plane wave. The accidental BIC is due to the singularity of the topological vortex of the far field radiative polarization, which ensures zero magnitude of radiation \cite{bicVortex}. By tuning the structural or dynamical parameter, the BIC can be transferred into quasi-BIC with ultra-high Q factor \cite{quasiBIC01,quasiBIC02,quasiBIC03,quasiBIC04,quasiBIC05,quasiBIC06,quasiBIC07,quasiBIC08,quasiBIC09,quasiBIC10,quasiBIC11,quasiBIC12,quasiBIC13,quasiBIC14,quasiBIC15,quasiBIC16,quasiBIC17,quasiBIC18,quasiBIC19,quasiBIC20,quasiBIC21}. Recently, momentum mismatch quasi-BICs have been proposed, which are due to folding the band structure of the guided resonance above the light cone into the first Brillouin zone \cite{bicZF}. The band folding is induced by the asymmetric parameter of the compound grating. The corresponding quasi-BIC have ultra-large quality factor, so that the GH shift can be highly enhanced \cite{bicGH1}. The quasi-BICs are off the $\Gamma$ point, so that the group velocity is nonzero.

For optical grating consisted of magneto-optical materials, such as YIG \cite{magOptYIG}, the band structure can be tuned by changing the magnetization of the materials \cite{magOpt1}. Because the magneto-optical effect break the time-reversal symmetry, the band structure become non-symmetric about the wave number ($k_{x}\rightarrow-k_{x}$), so that nonreciprocal response of the optical systems can be induced \cite{magOptNonrec}. Because the GH shift of the optical grating is dependent on the band structure, varying schemes have been proposed to control the GH shift by tuning the external magnetic field \cite{magOptControl1,magOptControl2,magOptControl3}. In contrast, by measuring the GH shift, the strength of the magneto-optical effect can be resolved \cite{magOptControl4}.

Previous researches about the GH shift require oblique incident of the optical beam, which could make the application schemes complicated. For the optical grating that preserves time reversal symmetry, the band structures at the $\Gamma$ point have zero group velocity. Thus, a normally incident quasi-plane wave cannot excite a travelling resonant mode at the $\Gamma$ point, so that the GH shift under normally incident optical beam is zero. In this paper, we proposed the GH shift of normally incident optical beam, which is induced by the quasi-BIC at $\Gamma$ point of compound grating of magneto-optical materials. Because of the magneto-optical effect, the angular spectrum of reflected and transmitted plane waves with incident angle being near zero could have large phase gradient, which enhance the GH shift. We designed a compound grating of magneto-optical rods, whose band structure at the $\Gamma$ point has non-zero group velocity as well as large Q factor. Thus, normally incident optical beam could have sizable GH shift, which is highly enhanced by tuning the asymmetric parameter of the compound grating.

The remainder of this paper is organized as follows: In Sec. II, the structure of the compound grating of magneto-optical rods is introduced, and the band structure is studied. In Sec. III, the angular optical spectrum and GH shift is calculated. In Sec. IV, the GH shift is studied in real space. In Sec. V, the conclusions are presented.

\section{band structure}

\begin{figure}[tbp]
\scalebox{0.58}{\includegraphics{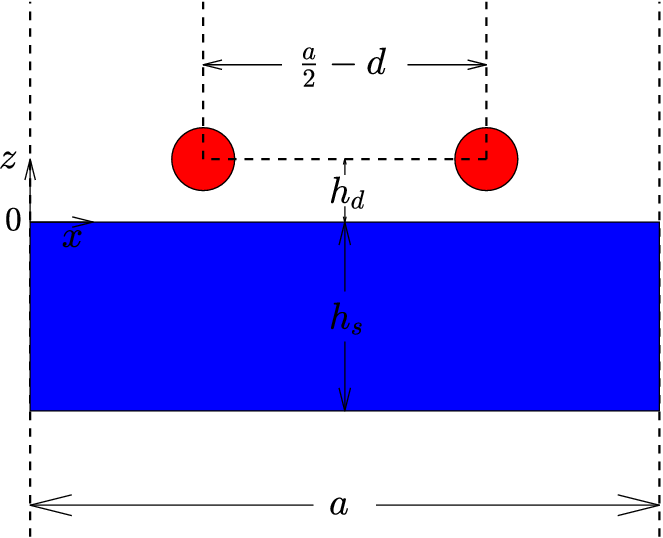}}
\caption{ The structure of the compound grating of YIG rods above the dielectric slab. The period of the grating is $a=3$ mm with two YIG rods in one period. The dielectric slab with refractive index being $3$ and thickness being $h_{s}=0.3a$ is indicated by the blue rectangular. The YIG rods are indicated by the red circles with radius bing $R=0.05a$. The center of the YIG rods are $h_{d}=0.1a$ above the dielectric slab. The distance between the two YIG rods in one period is $\frac{a}{2}-d$ with $d=0.05a$. The background medium is air. }
\label{figure_system}
\end{figure}

The structure of the compound grating of magneto-optical rods is given in Fig. \ref{figure_system}. A dielectric slab with thickness being $h_{s}$ and refractive index being 3 hosts waveguide modes. Periodic array of magneto-optical rod above the dielectric slab induce perturbation to the waveguide mode of the dielectric slab. In this paper, we consider YIG as magneto-optical material to demonstrate the physics. At microwave frequency range, the relative permittivity of the YIG is 15, and the relative permeability of the YIG is anisotropic, given as
\begin{equation}
\mu_{r}=\left[\begin{array}{ccc}
\mu_{d} & 0 & i\kappa  \\
0 & 1 &  0 \\
-i\kappa & 0 & \mu_{d} \\
\end{array}\right]
\end{equation}
With a 1600 Gauss magnetic field along $\hat{y}$ axis, the tensor elements at 4.28 GHz is $\mu_{d}=14$ and $\kappa=12.4$ \cite{YIGmu}. The time-reversal symmetry of the TM mode is broken. For a uniform array of magneto-optical rods with radius being $R=0.05a$, period being $a/2$, and the asymmetric parameter being $d=0$, the perturbation driven by the magneto-optical rods effectively fold the band of the waveguide mode of the dielectric slab into the first Brillouin zone, which is defined as $k_{x}\in[-2\pi/a,2\pi/a]$, and split the degeneration of the band structure at $k_{x}=\pm2\pi/a$. The compound grating is induced by laterally shifting the rods with even indices toward the rods with odd indices, i.e., nonzero asymmetric parameter or $d\ne0$. The size of period is double to be $a$. In one period, the distance between the two magneto-optical rods is $a/2-d$. When $d=0$, the array degrades into regular grating with period being $a/2$. When $d\ne0$, the array becomes compound grating with period being $a$, so that the first Brillouin zone become $k_{x}\in[-\pi/a,\pi/a]$. The band structures of the regular grating in the regime $\pi/a<|k_{x}|<2\pi/a$ are folded into the first Brillouin zone of the compound grating. With $d=0$, part of the band structure in the regime $\pi/a<|k_{x}|<2\pi/a$ is under the light cone, whose resonant modes have infinite Q factor. After the band folding induced by $d\ne0$, the resonant modes are folded into the regime above the light cone, and then the quality factor become finite. The Q factor of these modes exponentially approach infinite as the asymmetric parameter approach zero, i.e., $d\rightarrow0$. In the remaining part of the paper, the numerical result of the optical field are calculated by the Finite Element Method (FEM) in COMSOL. For the compound grating structure in Fig. \ref{figure_system}, Bloch periodic boundary conditions are applied to the left and right boundaries; perfectly matched layers (PMLs) are applied in the top and bottom boundaries to simulate the radiative background. For the band structure, eigenvalue problem of the Helmholtz wave equation of $E_{y}$ is solved. For the scattering problem, total-field scatter-field interfaces is applied in the region above grating to simulate the incident plane wave with small incident angle.

\begin{figure}[tbp]
\scalebox{0.58}{\includegraphics{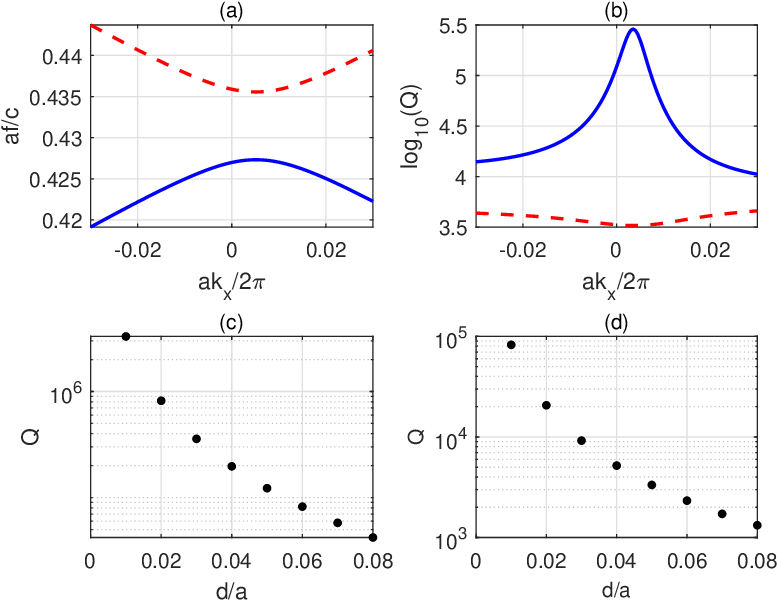}}
\caption{ (a) The band structure of the first and second TM modes of the component grating are plotted as blue (solid) and red (dashed) lines, respectively. (b) The Q factor of the corresponding bands are plotted. In panels (a,b), $d=0.05a$. The Q factor of the first and second TM modes at the $\Gamma$ point versus the asymmetric geometric parameter $d$ are plotted in (c) and (d), respectively. }
\label{figure_band}
\end{figure}

\begin{figure}[tbp]
\scalebox{0.58}{\includegraphics{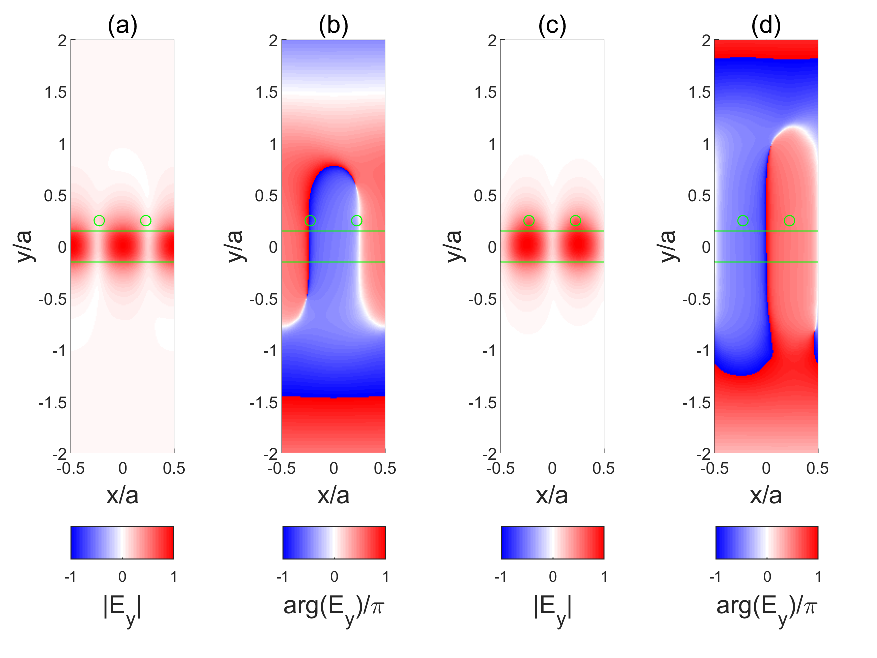}}
\caption{ The field pattern of the two resonant modes at the $\Gamma$ point in the band structure of Fig. \ref{figure_band}(c). The modes frequency being $af/c=0.435888$ and $0.426982$ are plotted in (a,b) and (c,d), respectively. Panels (a) and (c) are amplitude of the field pattern; panels (b) and (d) are phase of the field pattern. }
\label{figure_bandField}
\end{figure}

For a specific cases with $d=0.05a$, the band structures of the first and second TM mode near to the $\Gamma$ point are plotted in Fig. \ref{figure_band}(a). The Q factors of the corresponding bands are plotted in Fig. \ref{figure_band}(b). The two bands near to the $\Gamma$ point are folded from the edge of the first Brillouin zone of the system with $d=0$, i.e. near to $k_{x}=2\pi/a$, which are under the light cone. As $d$ change from $0$ to $0.08a$, the band structure hardly change, while the Q factors of the two modes at the $\Gamma$ point exponentially decrease, as shown in Fig. \ref{figure_band}(c) and (d). The first and second TM modes have larger and smaller Q factor. The field pattern of the two TM modes at the $\Gamma$ point are plotted in Fig. \ref{figure_bandField}. The majority of the optical energy is stored in the waveguide mode of the dielectric slab.
For the second TM mode at $\Gamma$ point, the two nodal lines of the field pattern cross the two magneto-optical rods, as shown in Fig. \ref{figure_bandField}(a) and (b). The overlap between the waveguide mode and the magneto-optical rods induces scattering with radiative loss. The field pattern is nearly even function of x coordinate, as shown by the phase of the field pattern in Fig. \ref{figure_bandField}(b), so that the mode is strongly coupled with the radiative plane wave. Thus, the Q factor is low. For the first TM mode at $\Gamma$ point, the peaks of the amplitude of the field pattern overlap with the two magneto-optical rods. The waveguide mode and the localized resonant mode in the two magneto-optical rods have strong coupling, as shown by the field pattern in Fig. \ref{figure_bandField}(c) and (d), where amplitude of the mode is large in the two magneto-optical rods as well as in the dielectric slab. The field pattern is nearly odd function of x coordinate, as shown by the phase of the field pattern in Fig. \ref{figure_bandField}(d), so that the mode is weakly coupled with the radiative plane wave. Thus, the Q factor is high.

\section{angular spectrum}

\begin{figure}[tbp]
\scalebox{0.58}{\includegraphics{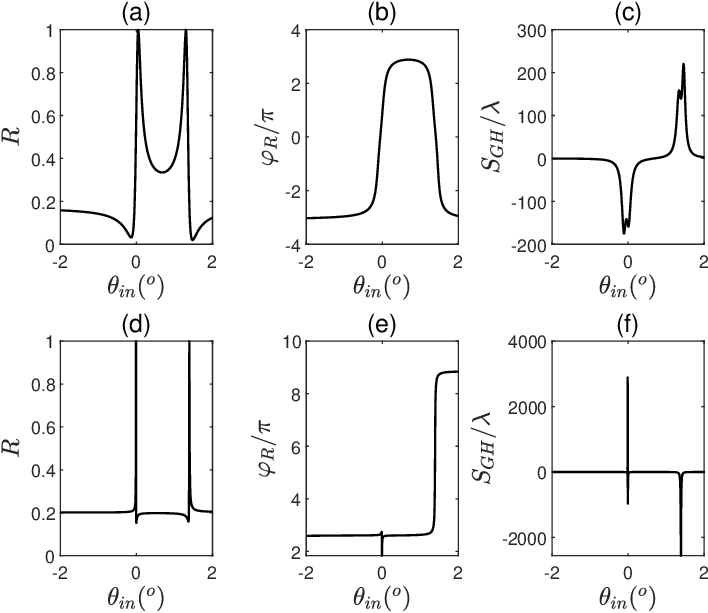}}
\caption{ Reflectance angular spectra near to normal incidence with frequency being $af/c=0.435888$ and $0.426982$ are plotted in (a) and (d), respectively. Corresponding reflection-phase angular spectra are plotted in (b) and (e), respectively. Corresponding GH shift angular spectra are plotted in (c) and (f), respectively. The structural parameters are the same as those in Fig. \ref{figure_band}(a,b). }
\label{figure_angle}
\end{figure}

\begin{figure}[tbp]
\scalebox{0.5}{\includegraphics{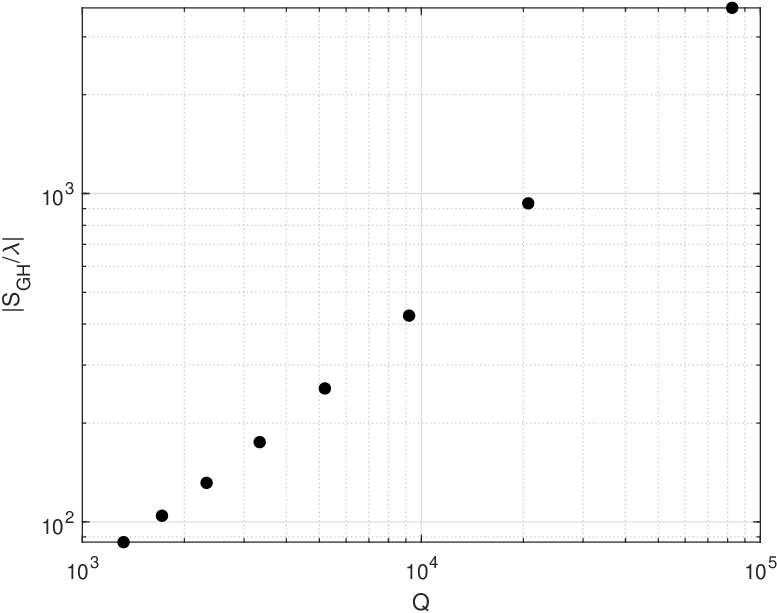}}
\caption{ Maximum magnitude of the negative GH shift corresponding to the second TM mode at $\Gamma$ point versus Q factor of the second TM mode at $\Gamma$ point. }
\label{figure_SGH_Q}
\end{figure}

The angular spectrum of reflectance with the frequency of the incident plane wave being equal to the frequency of the second and first TM modes at the $\Gamma$ point of the band structure in Fig. \ref{figure_band}(a) are plotted in Fig. \ref{figure_angle}(a) and (d), respectively. Because the isofrequency contours of the incident field cross the band structures of the first or second TM modes twice, there are two resonant peaks in each angular spectrum of the reflectance. The resonant peak that centers at zero incident angle corresponds to the excitation of the TM mode at the $\Gamma$ point. The full width at half maximum (FWHM) of the resonant peaks at zero incident angle of the second and first TM modes are $0.235^{o}$ and $0.025^{o}$, respectively. The angular spectrum of the phases of the reflected plane wave for the system in Fig. \ref{figure_angle}(a) and (d) are plotted in Fig. \ref{figure_angle}(b) and (e), respectively. Within each resonant peak, the phase dramatically change versus the incident angle. According to the stationary phase theory, the GH shift of an incident quasi-plane wave (Gaussian beam with beam width being infinitely large) is given as \cite{stationaryPt48}
\begin{equation}
S_{GH}=-\frac{\lambda}{2\pi}\frac{\partial\varphi_{R}}{\partial\theta_{in}}
\end{equation}
with $\varphi_{R}$ being the phase of the reflected plane wave, $\theta_{in}$ being the incident angle. The theoretical value of the GH shift are plotted in Fig. \ref{figure_angle}(c) and (f), for the system in Fig. \ref{figure_angle}(a) and (d), respectively. Because the first TM mode has larger Q factor, the corresponding angular spectrum of reflectance has narrower line width, and larger magnitude of GH shift. The sign of the GH shift switch between positive and negative within the line-width of the resonant peak, as shown in Fig. \ref{figure_angle}(f), so that the GH shift of a normally incident Gaussian beam with finite divergence angle would have both positive and negative GH shift. By contrast, for the GH shift corresponding to the excitation of the second TM mode at $\Gamma$ point, the sign is always negative within the line-width of the resonant peak, as shown in Fig. \ref{figure_angle}(c). As $d$ decreases, the Q factor of the second TM mode exponentially increases, which induces exponential increase of the maximum magnitude of the negative GH shift within the corresponding resonant peak, as shown in Fig. \ref{figure_SGH_Q}.

\section{Goos-H$\ddot{a}$nchen shift of normally incident Gaussian beam}

\begin{figure}[tbp]
\scalebox{0.58}{\includegraphics{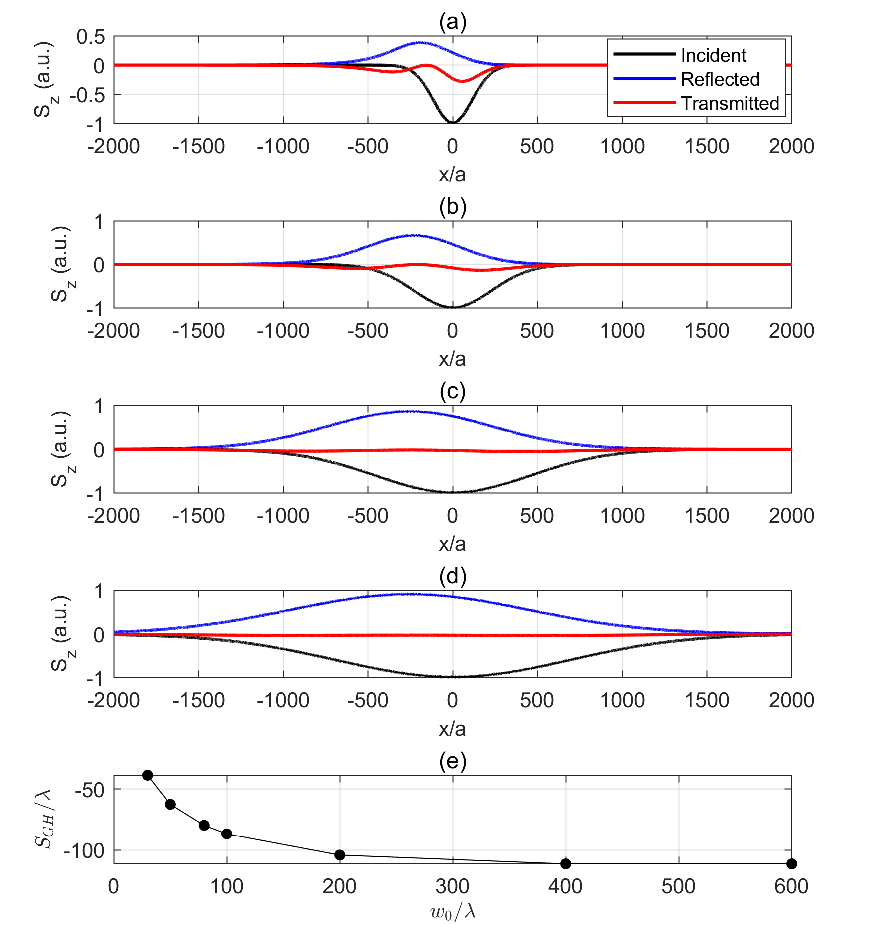}}
\caption{ In the presence of a normally incident Gaussian beam with $af_{Inc}/c=0.435888$, the time-averaged z-component Poynting vector of the incident and the reflected fields versus x coordinate at $z=5a$ are plotted as black and blue lines, respectively; and that of the transmitted field is plotted as red line. The beam width is $w_{0}/\lambda=100$, $200$, $400$, and $600$ in (a-d), respectively. (e) The GH shift obtained by fitting the reflected field versus the beam width of the incident Gaussian beam is plotted. The structural parameters are the same as those in Fig. \ref{figure_band}(a,b). }
\label{figure_GaussianB1}
\end{figure}

The scattering of normally incident Gaussian beam with finite beam width by the compound grating of the magneto-optical rods is numerically studied. For a normally incident Gaussian beam with fixed frequency and finite beam width, the field profile in the x-z plane is given as
\begin{equation}
\mathbf{E}_{0}(x,z)=E_{Inc}\hat{y}\sqrt{\frac{w_{0}}{w(z)}}e^{-\frac{x^{2}}{w^{2}(z)}-ik_{0}z-ik_{0}\frac{zx^{2}}{2(z^{2}+z_{0}^{2})}-\frac{i}{2}\eta(z)} \label{gaussianBeam}
\end{equation}
, where $w(z)=w_{0}\sqrt{1+(z/z_{0})^{2}}$, $\eta(z)=\tan^{-1}(-z/z_{0})$, $z_{0}=\frac{k_{0}w_{0}^{2}}{2}$, $k_{0}=2\pi f_{Inc}/c$, with $w_{0}$ being the beam width at the beam waist, $f_{Inc}$ being the frequency of the incident field. The focus of the Gaussian beam is at the top surface of the dielectric slab. The wavelength of the incident beam is given as $\lambda=c/f_{Inc}$. Because the beam width is finite, the optical field at the surface of the grating can be expanded as a superposition of multiple plane waves with varying incident angular. The spectrum of the incident plane wave is given as $\Theta(\theta)=e^{-(\theta/\Delta\theta_{in})^{2}/2}$, with $\Delta\theta_{in}=\sin^{-1}[\lambda/(\sqrt{2}\pi w_{0})]$ being the divergence angle of the Gaussian beam. The scattering of the Gaussian beam can be decomposed into superposition of the scattered plane waves. Because the phase of the scattered plane waves is dependent on the incident angle, the superposition effectively generate the scattered optical beam with beam center being shift along the lateral direction. If $\Delta\theta_{in}$ is much smaller than the FWHM of the angular spectrum of reflectance, the incident Gaussian beam is completely reflected, and the GH shift is the same as those given by the stationary-phase theory. On the other hands, if $\Delta\theta_{in}$ is larger than the FWHM of the angular spectrum of reflectance, the incident Gaussian beam is partially reflected. The spectrum of the reflected beam is given by the multiplication of $\Theta(\theta)$ and the angular spectrum of the reflected coefficient. Because the theoretical GH shift within the angular spectrum of the reflectance is a peak function, the GH shift of the reflected beam is the angular average of the theoretical GH shift within the angular spectrum, which is smaller than the GH shift at the peak. As a result, the GH shift of the incident Gaussian beam is dependent on $\Delta\theta_{in}$, or the beam width. The scattering of the normally incident Gaussian beam was calculated by FEM with total-field scatter-field interface at the top boundary of the computational domain, which simulates the normal incidence of the Gaussian beam. The time-averaged z-component Poynting vectors, designated as $S_{z}$, at the observation plane above and below the grating are plotted to visualize the GH shift.

\begin{figure}[tbp]
\scalebox{0.58}{\includegraphics{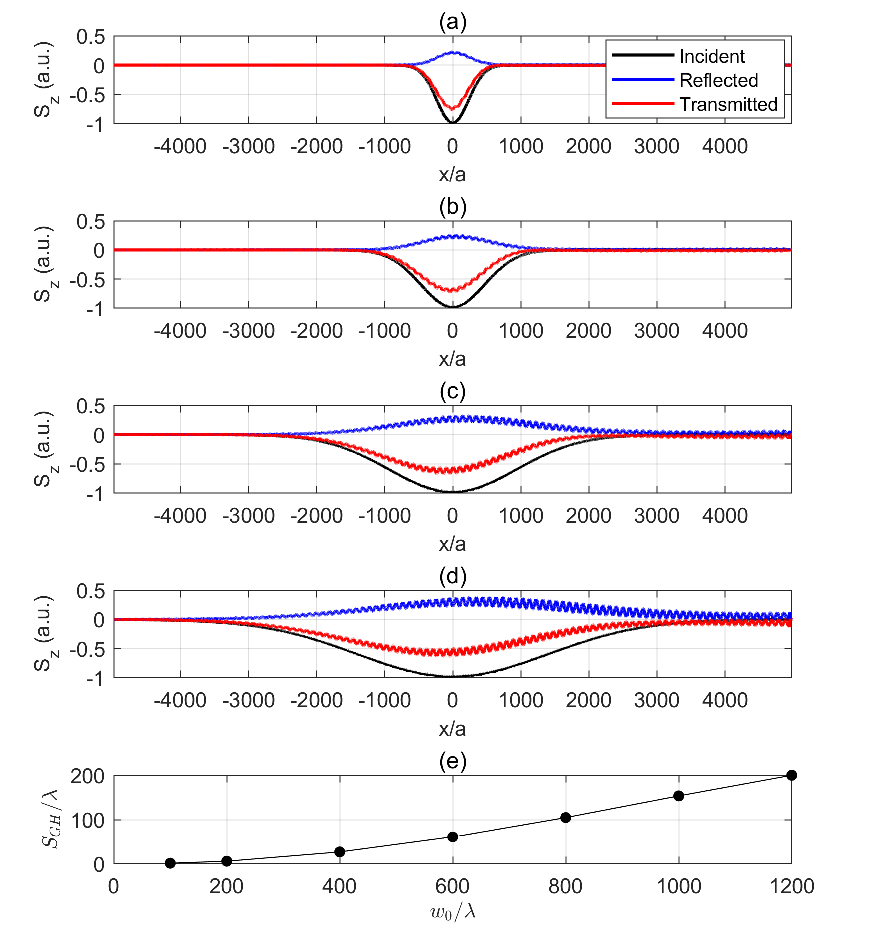}}
\caption{ The same as Fig. \ref{figure_GaussianB1}, except that $af_{Inc}/c=0.426982$ and $w_{0}/\lambda=200$, $400$, $800$, and $1200$ in (a-d).  }
\label{figure_GaussianB2}
\end{figure}

The scattering of normally incident Gaussian beam by component grating with $10000$ periods, and the structural parameters being the same as those in Fig. \ref{figure_band}(a,b) are plotted in Fig. \ref{figure_GaussianB1} and Fig. \ref{figure_GaussianB2}. As $f_{Inc}$ being equal to the second TM mode in Fig. \ref{figure_band}(a,b), the numerical results are shown in Fig. \ref{figure_GaussianB1}. For the incident Gaussian beam with $w_{0}/\lambda=100$, the divergence angle is $\Delta\theta_{in}=0.129^{o}$, which is slightly smaller than the FWHM of the reflectance peak. Thus, large portion of the incident energy is reflected, and a small portion of the incident energy is transmitted with both positive and negative GH shift, as shown in Fig. \ref{figure_GaussianB1}(a). As $w_{0}$ becomes larger, the divergence angle becomes smaller, and then more portion of the incident energy is reflected, as shown in Fig. \ref{figure_GaussianB1}(b-d). When $w_{0}$ reaches $600\lambda$, the reflectance of the incident Gaussian beam is nearly unity. The GH shift of the reflected beam can be estimated from the numerical results of the $S_{z}$ at the observation plane as
\begin{equation}
S_{GH}=\frac{\int_{-\infty}^{+\infty}{xS_{z}(x)dx}}{\int_{-\infty}^{+\infty}{S_{z}(x)dx}}\label{GHgaussian}
\end{equation}
The numerical results of $S_{GH}$ versus $w_{0}$ is plotted in Fig. \ref{figure_GaussianB1}(e). As $w_{0}$ being larger than $400\lambda$, $S_{GH}$ approaches $-111\lambda$, which is nearly equal to the GH shift given by the stationary-phase theory.

As $f_{Inc}$ being equal to the first TM mode in Fig. \ref{figure_band}(a,b), the numerical results are shown in Fig. \ref{figure_GaussianB2}. For the incident Gaussian beam with $w_{0}/\lambda=200$, the divergence angle is $\Delta\theta_{in}=0.065^{o}$, which is larger than the FWHM of the reflectance peak. Thus, only a small portion of the incident energy is reflected, as shown in Fig. \ref{figure_GaussianB2}(a). As $w_{0}$ becomes larger, the divergence angle becomes smaller, and then more portion of the incident energy is reflected, as shown in Fig. \ref{figure_GaussianB2}(b-d). When $w_{0}$ reaches $600\lambda$, the divergence angle is $\Delta\theta_{in}=0.011^{o}$, which is slightly smaller than the FWHM of the reflectance peak. Thus, large portion of energy is reflected, as shown in Fig. \ref{figure_GaussianB2}(d). Because the angular spectrum of the GH shift oscillates between position and negative value within the resonant peak of the reflectance, the reflected optical beam is consisted of interfere of two beams with positive and negative GH shift. Thus, $S_{z}$ of the reflected beam is oscillating versus $x$, as shown in Fig. \ref{figure_GaussianB2}(a-d). The numerical GH shift given by Eq. (\ref{GHgaussian}) versus $w_{0}$ is plotted in Fig. \ref{figure_GaussianB2}(e), which is the effective GH shift by averaging the positive and negative GH shift within the resonant peak. Thus, the value of the GH shift is much smaller than that given by the peak value of the GH shift given by the stationary-phase theory in Fig. \ref{figure_angle}. Because the group velocity of the first TM mode is positive, the averaged GH shift is positive.

\section{Conclusion}

In conclusion, component grating of magneto-optical rods induces quasi-BICs at the $\Gamma$ point with large Q factor and nonzero group velocity. Excitation of the quasi-BICs by normally incident Gaussian beam induces reflected beam with large GH shift. The magnitude of the GH shift is dependent on the Q factor of the quasi-BIC, as well as the beam width of the Gaussian beam.

\begin{acknowledgments}
This project is supported by  the Special Projects in Key Fields of Ordinary Universities in Guangdong Province(New Generation Information Technology, Grant No. 2023ZDZX1007), the Natural Science Foundation of Guangdong Province of China (Grant No. 2022A1515011578), and the Startup Grant at Guangdong Polytechnic Normal University (Grant No. 2021SDKYA117).
\end{acknowledgments}

\clearpage

\end{document}